\documentclass[conference]{IEEEtran}
\usepackage{booktabs}
\usepackage{cite}
\usepackage{amsmath,amssymb,amsfonts}
\usepackage{algorithmic}
\usepackage{graphicx}
\usepackage{textcomp}
\usepackage{xcolor}
\usepackage{multirow}
\usepackage{rotating}
\def\BibTeX{{\rm B\kern-.05em{\sc i\kern-.025em b}\kern-.08em
    T\kern-.1667em\lower.7ex\hbox{E}\kern-.125emX}}
\usepackage[activate]{microtype}
\begin{document}

\title{Audio Explanation Synthesis with Generative Foundation Models

}

\author{\IEEEauthorblockN{Alican Akman$^*$, Qiyang Sun$^*$, Bj\"{o}rn W. Schuller$^*$$^\dagger$}
\IEEEauthorblockA{\textit{$^*$GLAM, Imperial College London, UK} \\
\textit{$^\dagger$CHI, Technical University of Munich, Germany} \\
a.akman21@imperial.ac.uk}
}

\maketitle

\begin{abstract}
The increasing success of audio foundation models across various tasks has led to a growing need for improved interpretability to understand their intricate decision-making processes better. Existing methods primarily focus on explaining these models by attributing importance to elements within the input space based on their influence on the final decision. In this paper, we introduce a novel audio explanation method that capitalises on the generative capacity of audio foundation models. Our method leverages the intrinsic representational power of the embedding space within these models by integrating established feature attribution techniques to identify significant features in this space. The method then generates listenable audio explanations by prioritising the most important features. Through rigorous benchmarking against standard datasets, including keyword spotting and speech emotion recognition, our model demonstrates its efficacy in producing audio explanations.
\end{abstract}


\section{Introduction}

Generating explanations for large artificial intelligence (AI) models has been gaining importance as they are used in various domains such as audio processing and computer vision. Most existing explainable artificial intelligence (XAI) methods try to extract important features in the input space towards the model's final decision, that can be categorised into perturbation-based \cite{DBLP:journals/corr/RibeiroSG16, DBLP:journals/corr/abs-1806-07421, DBLP:journals/corr/ZeilerF13, DBLP:journals/corr/abs-1910-08485} and backpropagation-based \cite{10.1371/journal.pone.0130140, DBLP:journals/corr/SundararajanTY17, DBLP:journals/corr/ShrikumarGK17, DBLP:journals/corr/SelvarajuDVCPB16, DBLP:journals/corr/MontavonBBSM15} techniques \cite{DBLP:journals/corr/abs-1711-06104}. These methods aim to identify relevant input features, such as pixels for computer vision tasks and tokens for natural language processing tasks. On the other hand, providing audio explanations is a useful method due to its intuitiveness on audio-based tasks and higher expressiveness over other modalities in specific scenarios, such as where understanding visual explanations needs expertise \cite{10.1145/3462244.3479879}. Aiming to generate listenable and interpretable audio explanations, \cite{9871291, parekh2022listen} exploit non-negative matrix factorisation (NMF) \cite{NIPS2000_f9d11525} to decompose audio into meaningful components.

Foundation models are extensively used in audio processing to achieve state-of-the-art performance on various tasks such as automatic speech recognition, keyword spotting, and speaker recognition \cite{DBLP:journals/corr/abs-2104-01778, DBLP:journals/corr/abs-2106-07447, DBLP:journals/corr/abs-2110-13900, DBLP:journals/corr/abs-1910-05453, chi2021audio, ravanelli2020multitask, 9383605}. In addition to that, these models offer a generalised and meaningful embedding space due to their broad range of training data; some foundation models such as EnCodec \cite{defossez2022highfi} enable generation from this space. Although certain studies target to explain transformer-based foundation models by leveraging their attention mechanism and presenting attention weights as explanations \cite{DBLP:journals/corr/abs-1906-05714, yeh2023attentionviz, DBLP:journals/corr/abs-2005-00928}, they do not consider computing feature importance in the meaningful embedding space to understand model behaviour. Testing with Concept Activation Vectors (TCAV) \cite{kim2018interpretability} and Network Dissection \cite{bau2017network} focus on explaining a model's behaviour with provided concepts by exploiting the model internal representation. However, these methods require user-defined concepts without considering unleashing the learnt concepts which are already embedded in the latent space of a foundation model.

To address these issues, we propose a method which combines prominent feature attribution methods with foundation models to explain model behaviour in audio processing tasks. We first exploit an audio foundation model as an encoder, and train an additional model on this backbone depending on the type of the downstream task. To understand the behaviour of the final model on the task, we analyse important features for a decision in the latent space using a feature attribution method. In the final step, we use the generative part of the foundation model to construct the relevant audio in the input space. We verify that our method can generate high-fidelity explanations through experiments that simulate removing relevant features and assess the original model's performance on these essential features. The main contributions of this study are as follows:
\begin{itemize}
 \item 
We propose a novel audio explanation method that integrates common feature attribution methods into the latent space of foundation models. Our approach takes advantage of this meaningful space for creating understandable explanations without mapping feature relevance to the input space where individual features are difficult to interpret like audio frequencies.
 \item 
Our method leverages the generative capacity of foundation models from latent space to produce meaningful audio explanations in the input space. In this way, our method generates audio explanations in a listenable format that are interpretable to the end-user.
 \item 
We evaluate our model on keyword spotting and speech emotion recognition tasks. We show that while our method provides high-fidelity explanations, it captures meaningful high-level audio components for the investigated tasks.

\end{itemize}
\begin{figure}[t]
  \centering
  \includegraphics[width=\columnwidth, height=0.80\columnwidth]{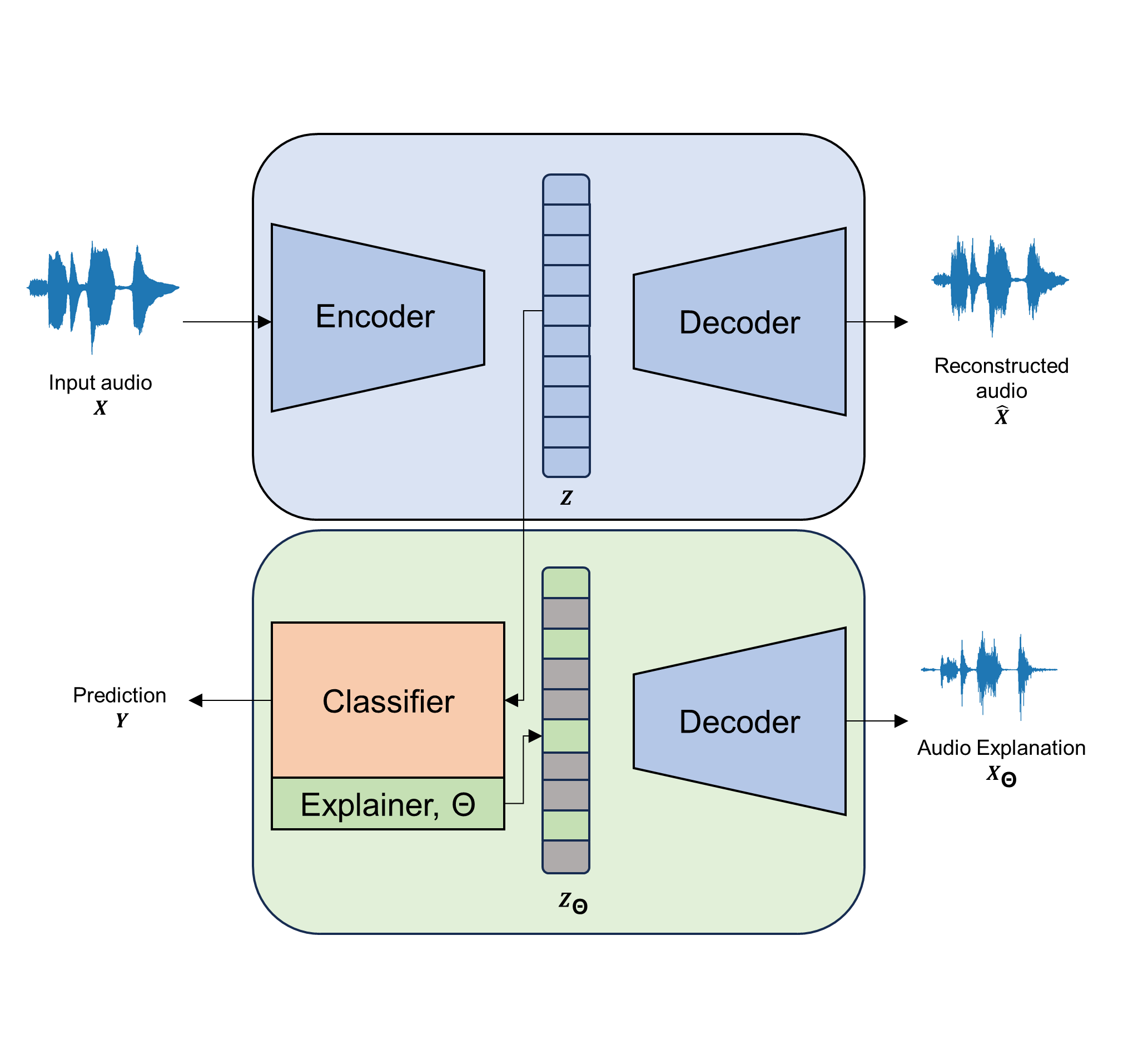}
  \caption{An overview of our method: The top row depicts the role of a foundation model with autoencoder architecture. The bottom row shows the process of explaining a task-specific classifier model including finding important features in the latent space and generating audio explanations based on these features.}
  \label{fig:res}
\end{figure}

\section{Related Work}

Adapting existing feature attribution methods to understand audio model predictions is a common practice. \cite{becker2019interpreting} explores the interpretability of deep audio models through the utilisation of layer-wise relevance propagation (LRP) \cite{10.1371/journal.pone.0130140, DBLP:journals/corr/BinderMBMS16}, a technique that computes relevance scores for each neuron in a deep neural network by recursively propagating relevance scores from the output. They examine the correlation between feature relevance scores and fundamental concepts like phonemes and distinct frequency ranges in classification tasks related to spoken digits and speaker gender. In \cite{frommholz2023xaibased}, the authors use DFT-LRP \cite{vielhaben2023explainable}, a recently introduced variant of LRP integrating Fourier transformation, to explain audio event detection models with different architectures. They evaluate the importance of individual time-frequency components regarding the predicted classes of the models. However, there is still room for enhancement in interpreting the feature importance maps provided by these methods.

Decomposing an audio input into meaningful components offers a practical approach for identifying the audio elements pertinent to XAI. CoughLIME \cite{9871291} extends the LIME method to explain audio processing models tailored specifically for cough data. A critical aspect of CoughLIME, distinguishing it from applying standard LIME to audio spectrograms, involves decomposing the input audio into interpretable components using NMF. The authors in \cite{parekh2022listen} introduce an interpreter network built from scratch, incorporating NMF as an audio decomposition technique. Their network is trained to develop surrogate models that replicate the output of the original classifier and generate temporal activations of pre-learnt NMF components. However, these methods primarily focus on computing relevance in the input space without delving into the intermediate representations within a deep model.

\section{Methodology}

This section elaborates on the design of our method to generate meaningful audio explanations. We begin by introducing the general structure of audio foundation models and their usage for downstream tasks. Following this, we provide a detailed description of our explainer system, which involves assigning importance in the embedding space of a foundation model. Lastly, we outline the steps for producing meaningful audio explanations using our approach. We present an overview of our system in Figure \ref{fig:res}.

\subsection{Audio Foundation Models}

Audio foundation models are typically pre-trained on large datasets of audio samples to learn patterns from the audio signals, which can then be fine-tuned on specific tasks. They exploit self-supervised learning strategies to discover general representations from large-scale data without requiring expensive labels. To learn these representations which reflect meaningful patterns in audio signals, they build a high-level embedding space using deep learning frameworks such as autoencoders. This framework uses an encoder-decoder pair to project the audio input into the embedding space and then reconstruct it. In this paper, we focus on audio foundation models using autoencoder architecture to be able to generate listenable explanations using its decoder component. The standard autoencoder architecture can be formulated as follows:

\begin{equation} \label{eq:FM}
Z = Encoder(X) \in{\mathbb{R}^{T \times L}} \quad ;\quad \hat{X} = Decoder(Z), 
\end{equation}

where $X\in{[-1,1]^{D \times F_s}}$ represents an audio signal input of duration $D$ with sample rate $F_s$, $Z$ represents the latent vector with $T$ denoting the number of audio frames after down-sampling in the encoder, $L$ the feature dimension of the encoder, and $\hat{X}$ represents the reconstructed audio signal.

\subsection{System Design}

We aim to understand the important audio features for a model decision by leveraging the high-level embedding space of foundation models. For this purpose, we target explaining foundation model-based audio models trained on specific audio tasks such as audio classification. Thus, our system starts with finetuning a foundation model with an autoencoder style framework on a desired task. To maintain the learnt representation space during finetuning, we freeze the weights of the encoder part and only update the additional task-specific model part such as a classification head. Then, our framework uses feature attribution methods to determine the most relevant features in the latent space for a model decision. Without backpropagating feature attribution computation to the audio input space, our method learns the important high-level components in the latent space, which is not restricted with the dimensions of the input space. We formulate our feature attribution method in the latent space as follows:

\begin{equation} \label{eq:att}
att = \Theta(Classifier(Z)) \in{\mathbb{R}^{T \times L}},  
\end{equation}

where $\Theta$ represents our explainer module that computes feature attribution, $att$, in the latent space. Note that the classifier module is designed to process the latent representation of an audio input extracted by the fixed encoder module.

\begin{table}[t]
\centering
\caption{Fidelity results by measuring explanation classification agreement on the Speech Commands and TESS dataset, with mean and standard deviation over five runs.}
\resizebox{\columnwidth}{!}{
\begin{tabular}{ c c c c c c } 
 \toprule
  & & \multicolumn{2}{c}{Latent Space} &  \multicolumn{2}{c}{Input Space} \\ 
 \cmidrule(lr){3-4}
 \cmidrule(lr){5-6}
 & Ratio, $\alpha$ & Our Method & Random & IG & Random \\ 
  \hline
 \multirow{5}{4em}{Speech \\ Commands} & $0.1$ & 79.1 $\pm$ 0.0 & 6.6 $\pm$ 1.1 & 18.6 $\pm$ 0.0 & 7.9 $\pm$ 0.5 \\
 & $0.2$ & 87.6 $\pm$ 0.0 & 16.5 $\pm$ 1.8 & 25.4 $\pm$ 0.0 & 10.2 $\pm$ 0.7 \\ 
 & $0.4$ & 91.6 $\pm$ 0.0 & 46.0 $\pm$ 1.9 & 27.3 $\pm$ 0.0 & 15.9 $\pm$ 0.8 \\ 
 & $0.6$ & 93.9 $\pm$ 0.0 & 69.6 $\pm$ 1.6 & 28.4 $\pm$ 0.0 & 23.0 $\pm$ 1.1 \\
 & $0.8$ & 97.1 $\pm$ 0.0 & 84.8 $\pm$ 1.3 & 31.2 $\pm$ 0.0 & 39.9 $\pm$ 1.9 \\ 
  \hline
 \multirow{5}{4em}{TESS} & $0.1$ & 21.9 $\pm$ 0.0 & 15.5 $\pm$ 0.0 & 57.7 $\pm$ 0.0 & 52.8 $\pm$ 0.9 \\
 & $0.2$ & 55.7 $\pm$ 0.0 & 16.0 $\pm$ 0.9 & 59.8 $\pm$ 0.0 & 54.2 $\pm$ 0.4 \\ 
 & $0.4$ & 62.5 $\pm$ 0.0 & 28.3 $\pm$ 1.5 & 62.3 $\pm$ 0.0 & 55.0 $\pm$ 0.3 \\ 
 & $0.6$ & 73.6 $\pm$ 0.0 & 38.3 $\pm$ 0.7 & 63.2 $\pm$ 0.0 & 55.4 $\pm$ 0.8 \\
 & $0.8$ & 90.4 $\pm$ 0.0 & 74.1 $\pm$ 1.4 & 62.6 $\pm$ 0.0 & 60.6 $\pm$ 0.7 \\ 
 \bottomrule
\end{tabular}
}

\label{table:sc_inc}
\end{table}

\subsection{Explanation Generation}

To generate listenable audio explanations, our method first extracts the relevant latent vector based on the computed feature attributions in Equation \ref{eq:att}. While keeping the latent dimensions with high importance, it replaces less important dimensions with a base latent vector which is obtained by encoding a noise audio with appropriate length. It then uses the decoder part of the foundation model of interest to transform the relevant latent vector into audio explanations. The explanation generation can be written as:

\begin{equation} \label{eq:expgen}
X_\Theta = Decoder(Z_\Theta),  
\end{equation}

where $Z_\Theta$ represents the relevant latent vector for a specific prediction and $X_\Theta$ represents the audio explanation in the input space. Although our method generates an audio explanation in the input space, it goes beyond only selecting features in this space by the integration of meaningful latent space.

\section{Experiments}

We evaluated our method on two datasets, namely, Speech Commands \cite{speechcommandsv2}, and the Toronto Emotional Speech Set (TESS) \cite{SP2/E8H2MF_2020}, to assess its performance across keyword spotting and speech emotion recognition tasks. In this section, we provide implementation details of our method, followed by a comprehensive quantitative and qualitative evaluation. The implementation code and sample audio explanations is available on our project page\footnote{https://github.com/glam-imperial/AudioXgen}. 

\begin{table}[t]
\centering
\caption{Fidelity results by measuring accuracy drop over explanation removal on the Speech Commands and TESS dataset, with mean and standard deviation over five runs.}
\resizebox{\columnwidth}{!}{
\begin{tabular}{ c c c c c c } 
 \toprule
  & & \multicolumn{2}{c}{Latent Space} &  \multicolumn{2}{c}{Input Space} \\ 
 \cmidrule(lr){3-4}
 \cmidrule(lr){5-6}
 & Ratio, $\beta$ & Our Method & Random & IG & Random \\ 
  \hline

 \multirow{6}{4em}{Speech \\ Commands} & $0.01$ & 20.4 $\pm$ 0.0 & 85.0 $\pm$ 0.4 & 51.4 $\pm$ 0.0 & 76.3 $\pm$ 1.1 \\
 & $0.1$ & 1.0 $\pm$ 0.0 & 83.2 $\pm$ 1.1 & 22.6 $\pm$ 0.0 & 56.2 $\pm$ 0.7 \\
 & $0.2$ & 0.7 $\pm$ 0.0 & 78.8 $\pm$ 2.3 & 10.4 $\pm$ 0.0 & 37.7 $\pm$ 0.6 \\ 
 & $0.4$ & 0.6 $\pm$ 0.0 & 66.4 $\pm$ 1.1 & 8.0 $\pm$ 0.0 & 20.3 $\pm$ 1.1 \\ 
 & $0.6$ & 0.4 $\pm$ 0.0 & 43.8 $\pm$ 1.8 & 7.8 $\pm$ 0.0 & 12.5 $\pm$ 0.4 \\
 & $0.8$ & 0.4 $\pm$ 0.0 & 18.9 $\pm$ 2.3 & 7.4 $\pm$ 0.0 & 7.6 $\pm$ 0.8 \\ 
  \hline
  
 \multirow{6}{4em}{TESS} & $0.01$ & 64.6 $\pm$ 0.0 & 96.3 $\pm$ 0.2 & 85.4 $\pm$ 0.0 & 94.9 $\pm$ 0.4 \\
 & $0.1$ & 22.1 $\pm$ 0.0 & 90.6 $\pm$ 0.7 & 55.2 $\pm$ 0.0 & 70.3 $\pm$ 1.4 \\
 & $0.2$ & 17.3 $\pm$ 0.0 & 74.7 $\pm$ 1.8 & 43.6 $\pm$ 0.0 & 59.6 $\pm$ 0.8 \\ 
 & $0.4$ & 16.9 $\pm$ 0.0 & 38.9 $\pm$ 1.2 & 39.8 $\pm$ 0.0 & 54.5 $\pm$ 0.7 \\ 
 & $0.6$ & 16.3 $\pm$ 0.0 & 28.2 $\pm$ 1.1 & 39.6 $\pm$ 0.0 & 53.4 $\pm$ 0.4 \\
 & $0.8$ & 17.1 $\pm$ 0.0 & 15.4 $\pm$ 1.2 & 40.1 $\pm$ 0.0 & 55.2 $\pm$ 0.7 \\ 
  
 \bottomrule
\end{tabular}
}
\label{table:sc_drop}
\end{table}

\subsection{Implementation Details}

We choose the EnCodec neural audio codec \cite{defossez2022highfi} as our foundation model to learn the audio representations from the raw waveform, which is trained across diverse domains including general audio, speech, and music. EnCodec comprises two key components: an encoder that extracts features based on a convolutional neural network (CNN), and a decoder module that reconstructs the same audio. While our method leverages the encoder module to extract meaningful audio representations and assign importance based on the classification model, the decoder part allows it to map these features to the input space. In our experiments, we use the EnCodec version for 24\,kHz audio at 1.5\,kbps bandwidth. We also eliminate the quantisation part of the model to increase the accuracy on the classification model with higher dimensional representation.

As our classification model, we train a transformer-based classifier on top of the embeddings extracted by EnCodec. Note that we only train a base model without extra tuning. In the classifier architecture for keyword spotting on Speech Commands, we use 3 layers of transformer with an 8 head multi-head attention module. We employ a dropout probability of 0.1 and the dimension of the feed-forward network model is 512 for each transformer layer. For speech emotion recognition on TESS, we use a gated recurrent unit (GRU) based recurrent neural network with 2 layers and 128 hidden dimensions and employ a dropout of 0.2. Our classifiers achieve an accuracy of $85.4\%$ on the test set for Speech Commands and $96.4\%$ on the arranged test set for TESS. To arrange the TESS test set, we select random emotions from each spoken word using 0.2 split ratio and share the test indices on our project page for reproducibility. To compute the feature attribution in the latent space, we use Integrated Gradients (IG) \cite{DBLP:journals/corr/SundararajanTY17} which calculate the integral of the gradients of the model’s output along the straight line path from the baseline to the input.

\begin{figure}[t]
  \centering
  \includegraphics[width=0.83\linewidth]{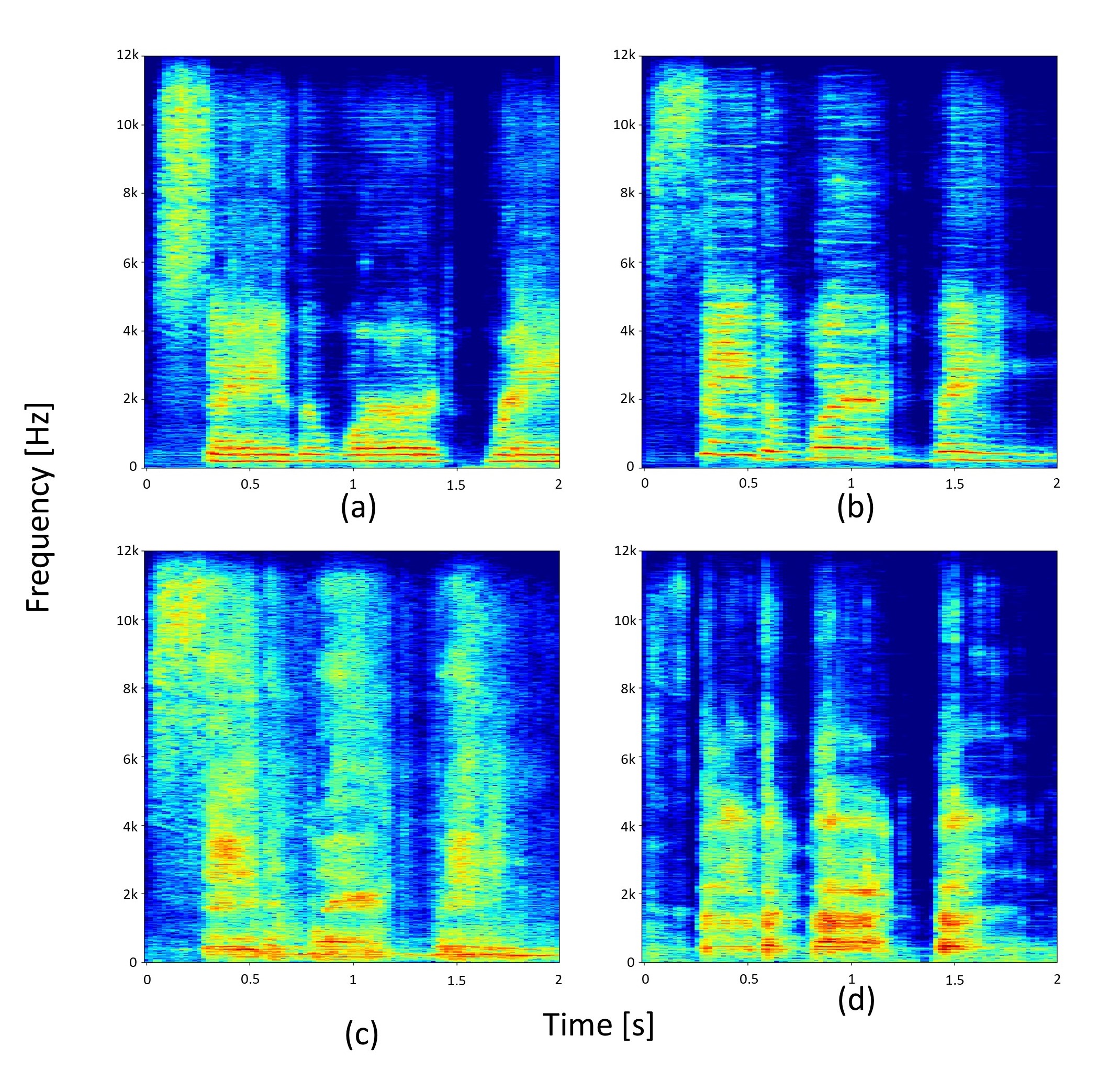}
  \caption{Sample spectrogram visualisations for the qualitative audio experiments: (a) Neutral audio of the word (``Rain"), (b) Happy audio of the word (``Rain"), (c) Explanation-removed audio from (b), (d) Explanation audio generated from (b).}
  \label{fig:qual}
\end{figure}

\subsection{Quantitative Evaluation}

We conduct fidelity experiments to measure how well the prediction of the underlying model and the generated explanation agree. Since our method investigates feature importance beyond the space of the original input features by integrating Encodec latent space, it is not possible to select important features in this space. Thus, our strategy involves selecting the latent dimensions with the highest relevance with respect to the IG algorithm by a ratio of $\alpha$. We set the remaining dimensions to a base value using a base latent vector which is obtained by encoding a noise audio with appropriate length. We compute the fidelity score as the fraction of samples where the predicted class for provided explanations in the latent space aligns with the classifier's prediction. We compare our methodology with three approaches: (1) We propose a baseline that randomly selects latent dimensions to obtain explanation embedding; (2) We select the most important features in the input audio space by a ratio of $\alpha$ using the IG method and generate the explanation embedding with the Encodec encoder -- here, eliminating the quantisation part of Encodec enables us to backpropagate the IG gradient calculation through the encoder safely; (3) We also implement the random feature selection strategy in the input audio space. The results in Table \ref{table:sc_inc} show that our method can generate explanations with higher fidelity compared to the two baselines (1) and (3) for both datasets. It also outperforms the standard IG method (2) which validates our approach on providing explanations leveraging a high-level embedding space.

In addition, we evaluate the fidelity of our method by analysing accuracy drop upon important feature removal to demonstrate that the generated explanations are relevant to the model's final decision. We measure the accuracy drop for different $\beta$ values which represent the ratio of most important features to be removed. Then, we compare our method with the other methods by following the same steps to produce the explanation embeddings. As shown in Table \ref{table:sc_drop}, feature removal based on our method leads to highest accuracy drop outperforming the other methods.

\begin{figure}[t]
  \centering
  \includegraphics[width=0.84\linewidth]{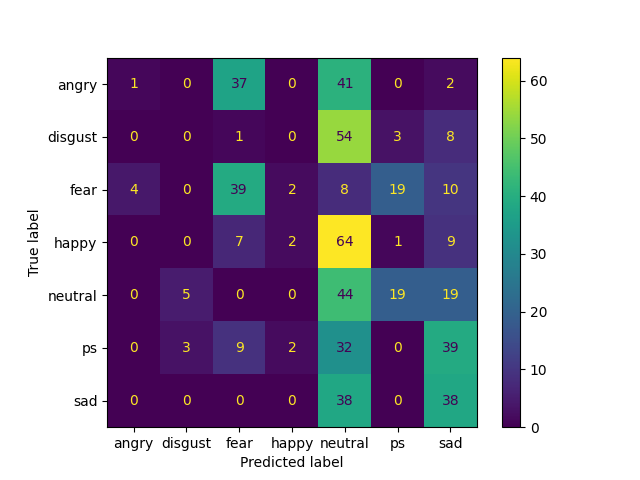}
  \caption{Confusion matrix for the classifier for TESS dataset after explanation removal by a ratio of $\beta = 0.1$.}
  \label{fig:confusion}
\end{figure}

\subsection{Qualitative Evaluation}

To evaluate the quality of generated explanations, we observe our model's behaviour on audios from the same spoken words and separate emotion classes on the TESS dataset. By following similar procedures with quantitative evaluation experiments, we select an audio sample and generate the audio explanation using our method by setting $\alpha = 0.2$ in the first experiment. In the second experiment setting, we removed the explanation from the original audio sample in the embedding space with the same ratio. We then generate the irrelevant audio part by using our framework. Ideally, we expect that while the generated audio in the first experiment can represent the emotion in the original audio sample, the irrelevant audio in the second experiment represents a neutral state of the same word spoken. In Figure \ref{fig:qual}, we present an example of audio from the class ``Happy" to conduct these experiments. We also present the audio of the same word from the class ``Neutral" to enable comparison for explanation-removed audio. We observe that while the generated explanation can represent the emotion in the original audio, explanation-removed audio looks more similar to the neutral version of the spoken word. We use spectrogram representation to increase the interpretability of the visuals.

We also investigate the classifier model behaviour for the TESS dataset after explanation removal by a ratio of $\beta = 0.1$ using our method. In Figure \ref{fig:confusion}, we present the confusion matrix of the classifier for each class in TESS dataset. The results show that majority of the audios are classified as ``Neutral" when explanation is removed by a small ratio which shows our explanation generation method focuses on emotions.

\section{Conclusion}

In this paper, we presented a novel audio explanation method which targets audio-processing foundation models. Unlike existing feature attribution methods which assign importance in the input space, our method integrates the latent space of a generative foundation model to generate meaningful and listenable explanations. The experiments demonstrated that our method delivers high-fidelity explanations, effectively capturing meaningful audio components pertinent to the specific task. Our work enlightens the way to promising research in interpreting state-of-the-art audio models as well as encompassing their application for model debugging and justification. An extension of our work could involve using rapidly growing audio generative AI models to produce higher quality audio explanations.

\clearpage
\bibliographystyle{IEEEtran2}
\bibliography{IEEEexample}

\begin{thebibliography}{10}
\providecommand{\url}[1]{#1}
\csname url@samestyle\endcsname
\providecommand{\newblock}{\relax}
\providecommand{\bibinfo}[2]{#2}
\providecommand{\BIBentrySTDinterwordspacing}{\spaceskip=0pt\relax}
\providecommand{\BIBentryALTinterwordstretchfactor}{4}
\providecommand{\BIBentryALTinterwordspacing}{\spaceskip=\fontdimen2\font plus
\BIBentryALTinterwordstretchfactor\fontdimen3\font minus \fontdimen4\font\relax}
\providecommand{\BIBforeignlanguage}[2]{{%
\expandafter\ifx\csname l@#1\endcsname\relax
\typeout{** WARNING: IEEEtran.bst: No hyphenation pattern has been}%
\typeout{** loaded for the language `#1'. Using the pattern for}%
\typeout{** the default language instead.}%
\else
\language=\csname l@#1\endcsname
\fi
#2}}
\providecommand{\BIBdecl}{\relax}
\BIBdecl

\bibitem{DBLP:journals/corr/RibeiroSG16}
\BIBentryALTinterwordspacing
M.~T. Ribeiro, S.~Singh, and C.~Guestrin, ``"why should {I} trust you?": Explaining the predictions of any classifier,'' \emph{CoRR}, vol. abs/1602.04938, 2016. [Online]. Available: \url{http://arxiv.org/abs/1602.04938}
\BIBentrySTDinterwordspacing

\bibitem{DBLP:journals/corr/abs-1806-07421}
\BIBentryALTinterwordspacing
V.~Petsiuk, A.~Das, and K.~Saenko, ``{RISE:} randomized input sampling for explanation of black-box models,'' \emph{CoRR}, vol. abs/1806.07421, 2018. [Online]. Available: \url{http://arxiv.org/abs/1806.07421}
\BIBentrySTDinterwordspacing

\bibitem{DBLP:journals/corr/ZeilerF13}
\BIBentryALTinterwordspacing
M.~D. Zeiler and R.~Fergus, ``Visualizing and understanding convolutional networks,'' \emph{CoRR}, vol. abs/1311.2901, 2013. [Online]. Available: \url{http://arxiv.org/abs/1311.2901}
\BIBentrySTDinterwordspacing

\bibitem{DBLP:journals/corr/abs-1910-08485}
\BIBentryALTinterwordspacing
R.~Fong, M.~Patrick, and A.~Vedaldi, ``Understanding deep networks via extremal perturbations and smooth masks,'' \emph{CoRR}, vol. abs/1910.08485, 2019. [Online]. Available: \url{http://arxiv.org/abs/1910.08485}
\BIBentrySTDinterwordspacing

\bibitem{10.1371/journal.pone.0130140}
\BIBentryALTinterwordspacing
S.~Bach, A.~Binder, G.~Montavon, F.~Klauschen, K.-R. Müller, and W.~Samek, ``On pixel-wise explanations for non-linear classifier decisions by layer-wise relevance propagation,'' \emph{PLOS ONE}, vol.~10, no.~7, pp. 1--46, 07 2015. [Online]. Available: \url{https://doi.org/10.1371/journal.pone.0130140}
\BIBentrySTDinterwordspacing

\bibitem{DBLP:journals/corr/SundararajanTY17}
\BIBentryALTinterwordspacing
M.~Sundararajan, A.~Taly, and Q.~Yan, ``Axiomatic attribution for deep networks,'' \emph{CoRR}, vol. abs/1703.01365, 2017. [Online]. Available: \url{http://arxiv.org/abs/1703.01365}
\BIBentrySTDinterwordspacing

\bibitem{DBLP:journals/corr/ShrikumarGK17}
\BIBentryALTinterwordspacing
A.~Shrikumar, P.~Greenside, and A.~Kundaje, ``Learning important features through propagating activation differences,'' \emph{CoRR}, vol. abs/1704.02685, 2017. [Online]. Available: \url{http://arxiv.org/abs/1704.02685}
\BIBentrySTDinterwordspacing

\bibitem{DBLP:journals/corr/SelvarajuDVCPB16}
\BIBentryALTinterwordspacing
R.~R. Selvaraju, A.~Das, R.~Vedantam, M.~Cogswell, D.~Parikh, and D.~Batra, ``Grad-cam: Why did you say that? visual explanations from deep networks via gradient-based localization,'' \emph{CoRR}, vol. abs/1610.02391, 2016. [Online]. Available: \url{http://arxiv.org/abs/1610.02391}
\BIBentrySTDinterwordspacing

\bibitem{DBLP:journals/corr/MontavonBBSM15}
\BIBentryALTinterwordspacing
G.~Montavon, S.~Bach, A.~Binder, W.~Samek, and K.~M{\"{u}}ller, ``Explaining nonlinear classification decisions with deep taylor decomposition,'' \emph{CoRR}, vol. abs/1512.02479, 2015. [Online]. Available: \url{http://arxiv.org/abs/1512.02479}
\BIBentrySTDinterwordspacing

\bibitem{DBLP:journals/corr/abs-1711-06104}
\BIBentryALTinterwordspacing
M.~Ancona, E.~Ceolini, A.~C. {\"{O}}ztireli, and M.~H. Gross, ``A unified view of gradient-based attribution methods for deep neural networks,'' \emph{CoRR}, vol. abs/1711.06104, 2017. [Online]. Available: \url{http://arxiv.org/abs/1711.06104}
\BIBentrySTDinterwordspacing

\bibitem{10.1145/3462244.3479879}
\BIBentryALTinterwordspacing
B.~W. Schuller, T.~Virtanen, M.~Riveiro, G.~Rizos, J.~Han, A.~Mesaros, and K.~Drossos, ``Towards sonification in multimodal and user-friendlyexplainable artificial intelligence,'' in \emph{Proceedings of the 2021 International Conference on Multimodal Interaction}, ser. ICMI '21.\hskip 1em plus 0.5em minus 0.4em\relax New York, NY, USA: Association for Computing Machinery, 2021, p. 788–792. [Online]. Available: \url{https://doi.org/10.1145/3462244.3479879}
\BIBentrySTDinterwordspacing

\bibitem{9871291}
A.~Wullenweber, A.~Akman, and B.~W. Schuller, ``Coughlime: Sonified explanations for the predictions of covid-19 cough classifiers,'' in \emph{2022 44th Annual International Conference of the IEEE Engineering in Medicine \& Biology Society (EMBC)}, 2022, pp. 1342--1345.

\bibitem{parekh2022listen}
J.~Parekh, S.~Parekh, P.~Mozharovskyi, F.~d'Alché Buc, and G.~Richard, ``Listen to interpret: Post-hoc interpretability for audio networks with nmf,'' 2022.

\bibitem{NIPS2000_f9d11525}
D.~Lee and H.~S. Seung, ``Algorithms for non-negative matrix factorization,'' in \emph{Advances in Neural Information Processing Systems}, T.~Leen, T.~Dietterich, and V.~Tresp, Eds., vol.~13.\hskip 1em plus 0.5em minus 0.4em\relax MIT Press, 2000.

\bibitem{DBLP:journals/corr/abs-2104-01778}
\BIBentryALTinterwordspacing
Y.~Gong, Y.~Chung, and J.~R. Glass, ``{AST:} audio spectrogram transformer,'' \emph{CoRR}, vol. abs/2104.01778, 2021. [Online]. Available: \url{https://arxiv.org/abs/2104.01778}
\BIBentrySTDinterwordspacing

\bibitem{DBLP:journals/corr/abs-2106-07447}
\BIBentryALTinterwordspacing
W.~Hsu, B.~Bolte, Y.~H. Tsai, K.~Lakhotia, R.~Salakhutdinov, and A.~Mohamed, ``Hubert: Self-supervised speech representation learning by masked prediction of hidden units,'' \emph{CoRR}, vol. abs/2106.07447, 2021. [Online]. Available: \url{https://arxiv.org/abs/2106.07447}
\BIBentrySTDinterwordspacing

\bibitem{DBLP:journals/corr/abs-2110-13900}
\BIBentryALTinterwordspacing
S.~Chen, C.~Wang, Z.~Chen, Y.~Wu, S.~Liu, Z.~Chen, J.~Li, N.~Kanda, T.~Yoshioka, X.~Xiao, J.~Wu, L.~Zhou, S.~Ren, Y.~Qian, Y.~Qian, J.~Wu, M.~Zeng, and F.~Wei, ``Wavlm: Large-scale self-supervised pre-training for full stack speech processing,'' \emph{CoRR}, vol. abs/2110.13900, 2021. [Online]. Available: \url{https://arxiv.org/abs/2110.13900}
\BIBentrySTDinterwordspacing

\bibitem{DBLP:journals/corr/abs-1910-05453}
\BIBentryALTinterwordspacing
A.~Baevski, S.~Schneider, and M.~Auli, ``vq-wav2vec: Self-supervised learning of discrete speech representations,'' \emph{CoRR}, vol. abs/1910.05453, 2019. [Online]. Available: \url{http://arxiv.org/abs/1910.05453}
\BIBentrySTDinterwordspacing

\bibitem{chi2021audio}
P.-H. Chi, P.-H. Chung, T.-H. Wu, C.-C. Hsieh, Y.-H. Chen, S.-W. Li, and H.~yi~Lee, ``Audio albert: A lite bert for self-supervised learning of audio representation,'' 2021.

\bibitem{ravanelli2020multitask}
M.~Ravanelli, J.~Zhong, S.~Pascual, P.~Swietojanski, J.~Monteiro, J.~Trmal, and Y.~Bengio, ``Multi-task self-supervised learning for robust speech recognition,'' 2020.

\bibitem{9383605}
E.~Kharitonov, M.~Rivière, G.~Synnaeve, L.~Wolf, P.-E. Mazaré, M.~Douze, and E.~Dupoux, ``Data augmenting contrastive learning of speech representations in the time domain,'' in \emph{2021 IEEE Spoken Language Technology Workshop (SLT)}, 2021, pp. 215--222.

\bibitem{defossez2022highfi}
A.~Défossez, J.~Copet, G.~Synnaeve, and Y.~Adi, ``High fidelity neural audio compression,'' \emph{arXiv preprint arXiv:2210.13438}, 2022.

\bibitem{DBLP:journals/corr/abs-1906-05714}
\BIBentryALTinterwordspacing
J.~Vig, ``A multiscale visualization of attention in the transformer model,'' \emph{CoRR}, vol. abs/1906.05714, 2019. [Online]. Available: \url{http://arxiv.org/abs/1906.05714}
\BIBentrySTDinterwordspacing

\bibitem{yeh2023attentionviz}
C.~Yeh, Y.~Chen, A.~Wu, C.~Chen, F.~Viégas, and M.~Wattenberg, ``Attentionviz: A global view of transformer attention,'' 2023.

\bibitem{DBLP:journals/corr/abs-2005-00928}
\BIBentryALTinterwordspacing
S.~Abnar and W.~H. Zuidema, ``Quantifying attention flow in transformers,'' \emph{CoRR}, vol. abs/2005.00928, 2020. [Online]. Available: \url{https://arxiv.org/abs/2005.00928}
\BIBentrySTDinterwordspacing

\bibitem{kim2018interpretability}
B.~Kim, M.~Wattenberg, J.~Gilmer, C.~Cai, J.~Wexler, F.~Viegas, and R.~Sayres, ``Interpretability beyond feature attribution: Quantitative testing with concept activation vectors (tcav),'' 2018.

\bibitem{bau2017network}
D.~Bau, B.~Zhou, A.~Khosla, A.~Oliva, and A.~Torralba, ``Network dissection: Quantifying interpretability of deep visual representations,'' 2017.

\bibitem{becker2019interpreting}
S.~Becker, M.~Ackermann, S.~Lapuschkin, K.-R. Müller, and W.~Samek, ``Interpreting and explaining deep neural networks for classification of audio signals,'' 2019.

\bibitem{DBLP:journals/corr/BinderMBMS16}
\BIBentryALTinterwordspacing
A.~Binder, G.~Montavon, S.~Bach, K.~M{\"{u}}ller, and W.~Samek, ``Layer-wise relevance propagation for neural networks with local renormalization layers,'' \emph{CoRR}, vol. abs/1604.00825, 2016. [Online]. Available: \url{http://arxiv.org/abs/1604.00825}
\BIBentrySTDinterwordspacing

\bibitem{frommholz2023xaibased}
A.~Frommholz, F.~Seipel, S.~Lapuschkin, W.~Samek, and J.~Vielhaben, ``Xai-based comparison of input representations for audio event classification,'' 2023.

\bibitem{vielhaben2023explainable}
J.~Vielhaben, S.~Lapuschkin, G.~Montavon, and W.~Samek, ``Explainable ai for time series via virtual inspection layers,'' 2023.

\bibitem{speechcommandsv2}
\BIBentryALTinterwordspacing
P.~{Warden}, ``{Speech Commands: A Dataset for Limited-Vocabulary Speech Recognition},'' \emph{ArXiv e-prints}, Apr. 2018. [Online]. Available: \url{https://arxiv.org/abs/1804.03209}
\BIBentrySTDinterwordspacing

\bibitem{SP2/E8H2MF_2020}
\BIBentryALTinterwordspacing
M.~K. Pichora-Fuller and K.~Dupuis, ``{Toronto emotional speech set (TESS)},'' 2020. [Online]. Available: \url{https://doi.org/10.5683/SP2/E8H2MF}
\BIBentrySTDinterwordspacing

\end{thebibliography}

\end{document}